\documentclass[a4paper,twocolumn,11pt,umpublished]{quantumarticle}
\pdfoutput=1
\usepackage[utf8]{inputenc}
\usepackage[english]{babel}
\usepackage[T1]{fontenc}
\usepackage{amsmath}
\usepackage{hyperref}

\usepackage{tikz}
\usepackage{lipsum}

\usepackage{graphicx}
\usepackage{amsmath,amssymb,amsfonts}

\usepackage{algorithm}
\usepackage{algpseudocode}

\usepackage{braket}

\begin{document}

\title{Differential Evolution VQE for Crypto-currency Arbitrage. Quantum Optimization with many local minima.}

\author{Gines Carrascal}
\affiliation{Department of Informatic Systems and Computation, Faculty of Informatics, Complutense University of Madrid, 28040 Madrid, Spain}
\orcid{0000-0001-7112-6696}
\affiliation{IBM Quantum, IBM Consulting España, 28830 Madrid, Spain}
\email{gines\_carrascal@es.ibm.com}

\author{Beatriz Roman}
\affiliation{Data Science, PwC España, 28830 Madrid, SPAIN}

\author{Guillermo Botella}
\affiliation{Department of Computer Architecture and Automation, Faculty of Informatics, Complutense University of Madrid, 28040 Madrid, Spain}
\orcid{0000-0002-0848-2636}

\author{Alberto del Barrio}
\affiliation{Department of Computer Architecture and Automation, Faculty of Informatics, Complutense University of Madrid, 28040 Madrid, Spain}
\orcid{0000-0002-6769-1200}

\maketitle

\onecolumn

\begin{abstract}
Crypto-currency markets are known to exhibit inefficiencies, which presents opportunities for profitable cyclic transactions or arbitrage, where one currency is traded for another in a way that results in a net gain without incurring any risk. Quantum computing has shown promise in financial applications, particularly in resolving optimization problems like arbitrage. In this paper, we introduce a differential evolution (DE) optimization algorithm for Variational Quantum Eigensolver (VQE) using Qiskit framework. We elucidate the application of crypto-currency arbitrage using different VQE optimizers. Our findings indicate that the proposed DE-based method effectively converges to the optimal solution in scenarios where other commonly used optimizers, such as COBYLA, struggle to find the global minimum. We further test this procedure's feasibility on IBM's real quantum machines up to 127 qubits. With a three-currency scenario, the algorithm converged in 417 steps over a 12-hour period on the "ibm\_geneva" machine. These results suggest the potential for achieving a quantum advantage in solving increasingly complex problems.
\end{abstract}

\section{Introduction}
\label{sec:introduction}

For millennia, commercial practice was based on a combination of barter and other means of exchange and, it was not until the introduction of coinage in the Greek regions at the late seventh century BC, that the first roots of arbitrage were born. Ancient arbitration was related to the trading of coins and ingots for profit, thus giving rise to the first desire to develop this practice \cite{poitras_origins_2021}.

Arbitrage could be defined as a financial strategy in which the investor obtains a profit from the difference in price of the same financial asset between different markets, this being a risk-free operation. The basic idea of this concept is to carry out complementary operations (buying and selling) taking advantage of market irregularities \cite{ross_return_1973}. The importance of simultaneity of transactions in arbitrage is crucial to avoid exposure to market risk or execution risk, the latter being the risk that prices change before the purchase and sale of the asset has been completed. This pursues that the arbitrage is opposite to the idea that the market is perfectly efficient and all the assets converge to the same price \cite{kabasinskas_key_2021}.

Due to their properties, financial products that are digitally traded are ideally suited for arbitrage. That is why, with the emergence of digital assets, the crypto-currency market has burst with great force. In this case and unlike traditional markets, they lack of a centralized exchange, so it is an immature market where there are periods with great arbitrage opportunities \cite{kabasinskas_key_2021}.

To acquire cryptocurrencies it is necessary to go to an exchange or cryptobank. There are currently more than six hundred exchanges worldwide, with \textit{Binance, FTX or Coinbase} being the best-valued exchanges with the highest trading volume to date \cite{coinmarketcap_top_2023}. Therefore, arbitrage opportunities emerge from the price difference between the different exchanges as it is an unregulated market. In this line, there are several types of cryptocurrency arbitrage depending on the number of assets traded and the different exchanges involved. If two cryptocurrencies are bought and sold between two different exchanges, we would be talking about parallel arbitrage. If three cryptocurrencies and two or even three different exchanges are involved, it would be a triangular arbitrage.
It is important to note that, in order to carry out these two types of arbitrage, it is necessary to have the cryptocurrencies to be traded in all the exchanges involved. The rebalancing technique, known as the action of sending cryptocurrencies between different exchanges, is usually subject to the payment of commissions. This is the reason why the arbritage intraexchange is one of the most used and therefore the one that allows the most operations to be made.  The process consists in transferring the capital from one cryptocurrency to another, always starting and ending in the same one thus being able to avoid commissions \cite{aldeano_moreno_estrategias_2020}. Throughout this paper we will focus on the latter type of arbitration.

This paper is organized as follows: 
Section \ref{sec:sota} contextualizes the ccurrent approaches to arbitrage throug quantum. 
Section \ref{sec:vqe} introduces the idea of quantum optimization using VQE. 
Section \ref{sec:arbitrage} explains the implications of arbitrage problem.
Section \ref{sec:quantum_arbitrage} describes the possibilities of using quantum computing for arbitrage. 
Section \ref{sec:global_de} introduces diferential evolution as well as its benefits and drawbacks.
Sections \ref{sec:arbitrage_3} introduces the model with 3 currencies, 
Sections \ref{sec:arbitrage_4} expand the model to 4 currencies, 
and Section \ref{sec:arbitrage_5} describes the quantum algorithms that can be applied to real data from 5 crypto-curencies, and details the considerations we take with real quantum computers.
Section \ref{sec:conclusion} contains our conclusions.

\section{State of the art}\label{sec:sota}
Given the multitude of opportunities offered by the cryptocurrency market \cite{makarov_trading_2020}, trading algorithms have been developed to exploit short-term arbitrage opportunities, thus making a profit from optimization algorithms \cite{kabasinskas_key_2021}. This is why quantum computing emerges as a complementary and more efficient alternative in certain processes with a greater speed and versatility unattainable by classical computing. 

The auge of machine learning research has also reached finance. Consequently, initial machine-learning-based statistical arbitrage strategies have emerged \cite{fischer_statistical_2019}. Also there are authors that ptopose mechanism to autoregulate the cryptocurrency market and compensate arbitrage opportunities \cite{krishnamachari_dynamic_2021, wang_optimal_2022}.

Especially in the banking sector, the importance of quantum computing in solving optimization algorithms has been demonstrated. In this sector, problems try to be simplified by reducing the number of possible solutions, which is why quantum computers can become very useful when the number of variables, and therefore the complexity of the problem, increases. Given the computational intensity of financial problems, increasingly more institutions are betting on the use of quantum technology  to solve arbitrage problems as well as portfolio optimization and price scoring. It is expected that the limits of quantum computing will be progressively reduced in order to take advantage of the full potential of this technology in the near future \cite{bova_commercial_2021}. 

Many industries are trying to develop quantum algorithms to address financial problems. In particular, companies such as \textit{IBM} and \textit{McKinsey} see optimization as one of the most in-demand use cases in the financial industry \cite{byrum_quantum_2021}. This is because most financial problems can be formulated as optimization problems, which is a particularly challenging task for classical computers. This is where the application of quantum algorithms makes it possible to easily solve such problems \cite{orus_quantum_2019}.

In the case of financial arbitrage, the problem can be formulated as a graph in which the different assets are the vertices. In 2007, Wanmei Soon and Heng-Qing Ye proposed a binary optimization model based on binary integer programming (BIP) by which the unboundedness problem of classical linear programming (LP) solving was solved. In this model, the authors established the existence of a feasible solution and a bounded objective function value \cite{soon_currency_2007}. Later in 2016, Rosenberg introduced several key ideas in arbitrage algorithms. First, the difference between detecting an arbitrage opportunity and detecting the optimal arbitrage solution, thus creating an alternative that not only finds the best solution, but those closest to it. To this end, he introduced quantum resolution by transforming the Wanmei Soon and Heng-Qing problem into a quadratic binary unconstrained binary optimization (QUBO) problem by rewriting the constraints as penalty terms. In addition, Rosenberg developed an alternative model that included backward arbitrage strategies, that is, the same asset (vertex) could be revisited \cite{orus_quantum_2019}.

To the best of our knowledge, these work is the firs aplying quantum optimization with a differential evolution approach to the arbitrage problem. There is a work from Zhuang et al. that proposes quantum algorithms for high-frequency statistical arbitrage trading, which is not exactly the same problem and in a different approach \cite{zhuang_quantum_2022}.  

In March 2017 Qiskit was launched, a software designed to create quantum algorithms, connect them to a back-end device and run them on simulators and real hardware \cite{qiskit_contributors_qiskit_2023}. In it, quadratic optimization models can be created using DocPLEX. IBM Decision Optimization CPLEX for python, known as DocPLEX, is an optimization software that solves quadratic models (QP), i.e. models with linear constraints and objective function with one or more quadratic terms. CPLEX can also solve convex problems efficiently \cite{ibm_solving_2021}. Once the quadratic problem is modeled, it is converted to QUBO to be solved applying quantum computing. 

To date, there is still a large gap between the resources available on current hardware and those demanded by some of the most relevant quantum applications. However, quantum research is advancing rapidly and increasingly more companies are betting on the research and use of quantum software. Big companies or new startups specialized in quantum computing in finance and economics, are working on the creation of new combinatorial optimization algorithms applicable to arbitrage with main banking institutions \cite{lee_quantum_2019}. 

\section{Variational Quantum Eigensolver (VQE)}\label{sec:vqe}

Despite the evolution of quantum software, many quantum optimization algorithms have hardware requirements that exceed the capacity of current quantum computers and may even fail to run. 

\begin{figure}[h]
  \centering
  \includegraphics[width=0.7\textwidth]{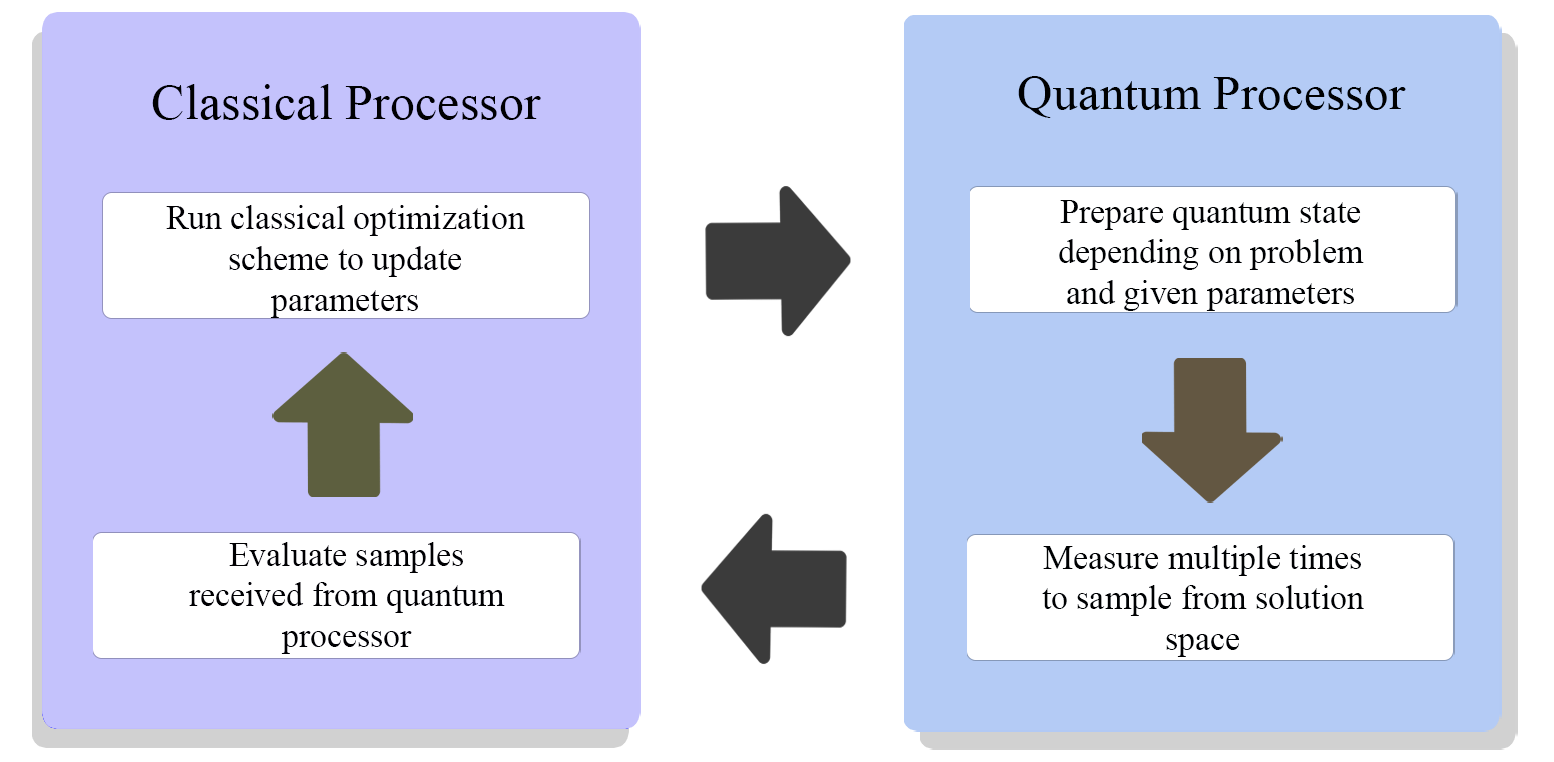}
  \caption{Hybrid quantum-classical algorithm.}
  \label{fig:hybrid_algorithm}
\end{figure}

That is why, in 2014 Peruzzo and McClean et al. \cite{peruzzo_variational_2014} developed VQE (Variational Quantum Eigensolver) a hybrid quantum-classical algorithm capable of finding variational solutions to problems that are not supportable by classical computers. Fig. \ref{fig:hybrid_algorithm} shows the processing cycle between a quantum and a classical processor. VQE has an advantage over classical algorithms in that a quantum processing unit can represent and store the exact wave function of the problem, which is extremely difficult for a classical computer \cite{mcclean_theory_2016}.

The VQE algorithm starts with a parameterized quantum circuit called ansatz and searches for the optimal parameter for this circuit using a classical optimizer. The ansatz is varied, via its set of parameters, by the optimizer, such that it works towards a state, as determined by the parameters applied to the ansatz, that will result in the minimum expectation value being measured of the input operator which is a Hamiltonian as (\ref{VQE}). The Ising Hamiltoninan, $H$, can be decomposed as a sum of Pauli terms and with its ground state (minimal energy state) corresponding to the optimal solution of original optimization problem.

Given $H$, with an unknown minimum eigenvalue $\lambda_{min}$, associated with an eigenstate $|\psi_{min}\rangle $, VQE provides an estimate $\lambda_\theta$ bounding $\lambda_{min}$ where $|\psi(\theta)\rangle$ is the eigenstate associated with $\lambda_\theta$. By applied the ansatz, $U(\theta)$, to some arbitrary starting state $|\psi\rangle$, VQE obtains an estimate $U(\theta)|\psi\rangle \equiv |\psi(\theta)\rangle$ on $|\psi_{min}\rangle$. This estimate is iteratively optimized by the classical optimizer changing the parameter $\theta$ minimizing the expectation value of $\langle\psi(\theta)| H |\psi(\theta)\rangle$ \cite{mcclean_theory_2016}.

\begin{equation} \label{VQE}
    \lambda_{min} \leq \lambda_\theta \equiv \langle\psi(\theta)| H |\psi(\theta)\rangle~.
\end{equation}

Using Qiskit, the Ising Hamiltonian can be obtained from a QUBO type problem created with DocPLEX. That is, a quadratic, binary and unconstrained problem. This aspect will be detailed in future sections.

\begin{figure}[h]
  \centering
  \includegraphics[width=\textwidth]{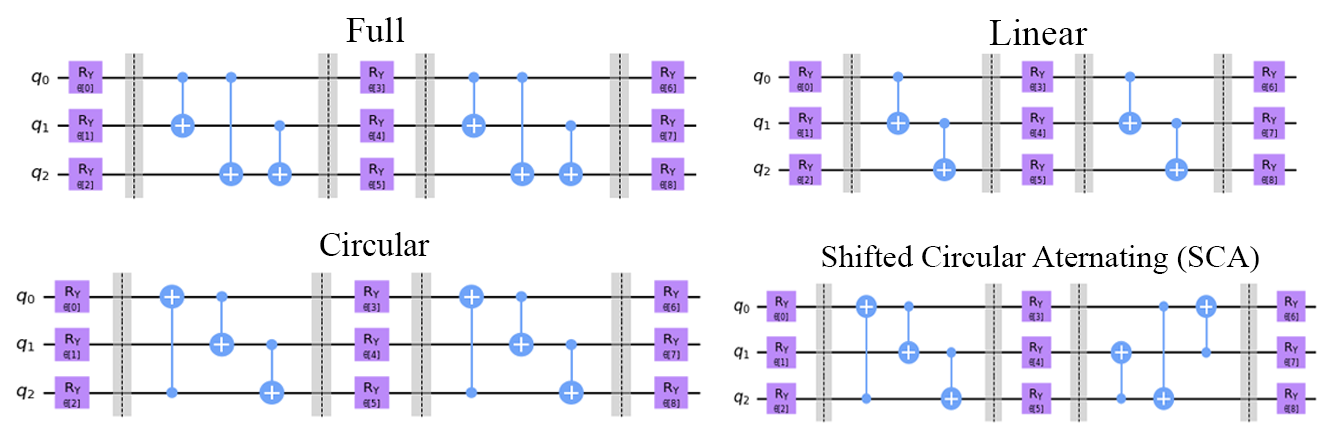}
  \caption{Types of entanglement on the Ansatz.}
  \label{fig:ansatz_types}
\end{figure}

In addition, it is possible to define the quantum circuit of the ansatz. There are several ways to create the ansatz, one of the most used is to create a heuristic circuit called Real Amplitudes. This type of circuit consists on alterning layers of gates Y rotations (purple gates in Fig. \ref{fig:ansatz_types}) and CNOT entanglements (blue gates in Fig. \ref{fig:ansatz_types}). Depending on the type of entanglement between the qubits we can create different ansatz as shown in Fig. \ref{fig:ansatz_types}.

It is also possible to set the number of repetitions, in this case, Fig. \ref{fig:ansatz_types} has two repetitions and they can be distinguished by the barriers of the circuit. Different ansatz can lead to different solutions in the optimization algorithm, so it is interesting to adjust these two hyper-parameters.

\section{Arbitrage as an optimization problem}\label{sec:arbitrage}

As mentioned, there are several types of crypto-currency arbitrage and, in this case, we will illustrate the problem of intra-exchange arbitrage in which $n$ currencies are involved and the exchange rate between these $n$ currencies is calculated. As we recall, in this type of arbitrage only a single exchange is involved. 

To illustrate the problem, the exchange prices of three crypto-assets will be simulated. This is because to solve the problem quantumly, it is necessary to connect to a quantum simulator in the cloud and therefore, the execution time is much higher than if we execute the problem on a real quantum computer, which would take milliseconds. That is why, first we will pose a small-scale problem with simulated data to test which optimizers work better and later, we will increase the number of crypto-assets and simulate a problem with real data.

\begin{figure}[h]
  \centering
  \includegraphics[width=0.7\textwidth]{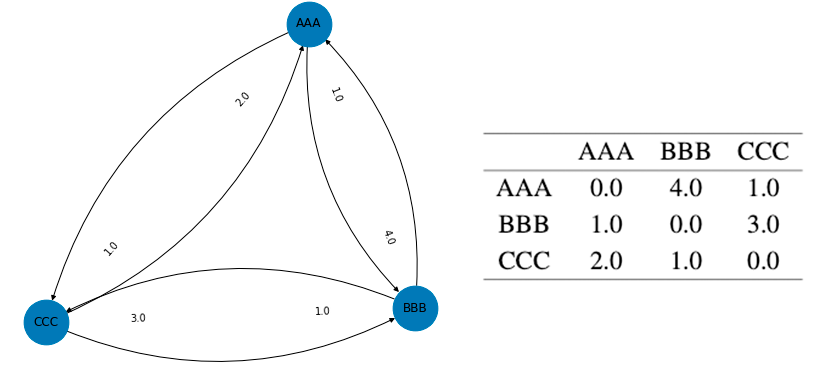}
  \caption{Exchange rates between three crypto-currencies and the transit price matrix.}
  \label{fig:exchange_rates}
\end{figure}

Originally, and in order to set up the model, the problem can be presented as a graph as shown in the Fig. \ref{fig:exchange_rates}. In this case we have three crypto-assets with a clear arbitrage opportunity.

\subsection{Initial formulation of the problem}

The first step in solving the arbitrage problem is to set up the optimization model. In this case, in (\ref{Model1}) an objective function is posed where the profit is maximized based on the exchange rates. We take logarithms of the elements of the transit matrix, $\Sigma\in\mathbb{R}^{nxn}$, and two constraints are established. The first establishes a closed arbitrage circle, i.e., it must start and end in the same currency, while the second establishes that each currency can only be traded once. Notice that $x \in \{0, 1\}^{n \times n}$ denotes the matrix of binary decision variables, which indicate which edge to pick ($x[i] = 1$) and which not to pick ($x[i] = 0$). In this problem we will assume that the commissions are included in the transit matrix. Remember that the model will be computationally created with DocPLEX for easy conversion to quantum mode. 

\begin{equation} \label{Model1}
\begin{aligned}
& \quad \max_{x\in\{0,1\}^{nxn}} x\Sigma~, \\
\textrm{s.t:} &\quad \sum_{i=1}^{n} x_{ij} = \sum_{j=1}^{n} x_{ij}~, \\
& \quad \sum_{i=1}^{n} x_{ij} \leq 1~. \\
\end{aligned}
\end{equation}

\begin{figure}[h]
  \centering
  \includegraphics[width=0.4\textwidth]{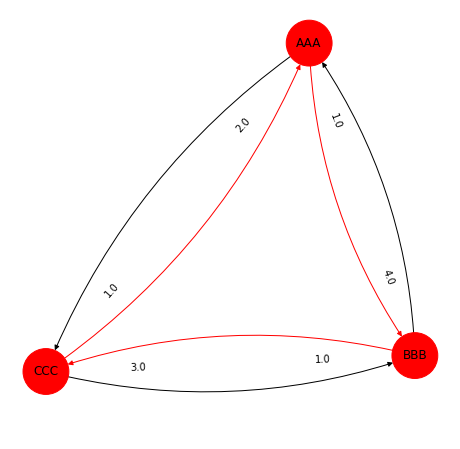}
  \caption{Solution of the arbitrage optimization problem: best arbitrage path.}
  \label{fig:classical_solution}
\end{figure}

Fig. \ref{fig:classical_solution} shows the result of running the optimization algorithm classically. 

In this case, the profit rate was 23, that is the difference between the initial value, and the value obtained after the loop.
The disadvantage of using only the classical method is that the arbitrage problem scales, that is, the classical algorithm calculates the result by "brute force" by trying all possible combinations, which is why if we add more crypto-assets to the problem and, therefore, to the graph, the problem would grow exponentially. For this reason, a quantum-classical method is necessary. 

\subsection{QUBO transformation}

\begin{figure}[h]
  \centering
  \includegraphics[width=\textwidth]{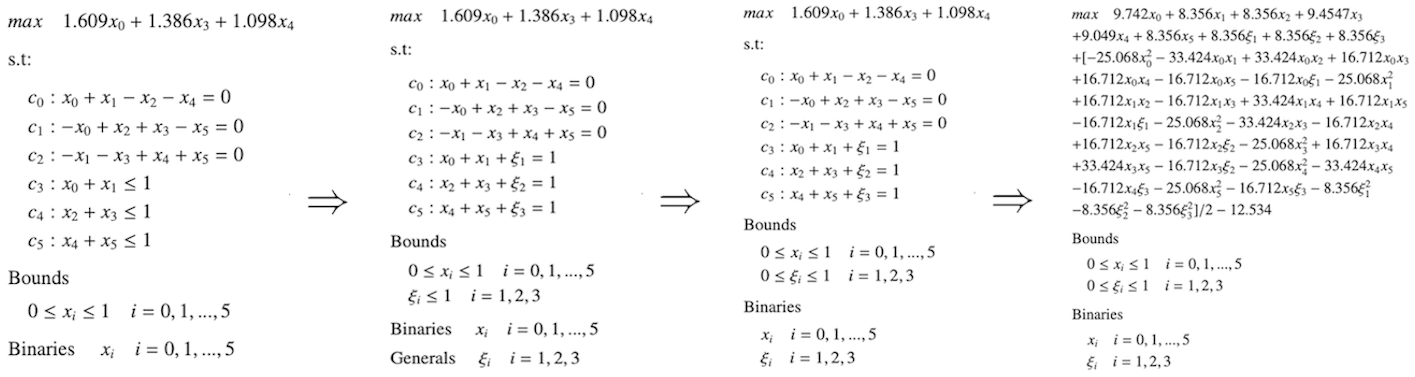}
  \caption{QUBO steps.}
  \label{fig:QUBO_steps}
\end{figure}

To apply the quantum method it is necessary to have the problem in QUBO form. If we remember, it is a binary, unconstrained and quadratic problem. The first step is to convert the inequality constraints into equality using slack variables, the signs of these variables depend on the inequality symbol. Successively, our model was originally a binary model, but when introducing the slack variables, it is necessary to convert these to binary as well. The last step is to convert the model to an unconstrained model. Up to this point, the model had equality constraints, so to remove them, they will be converted into additional quadratic penalty terms of the objective function. All in all, we would have a quadratic, binary, unconstrained problem just as we needed. These four steps are reflected in Fig. \ref{fig:QUBO_steps} where the evolution of the model until the QUBO format is achieved can be seen. To solve this problem we will need 9 qubits as we have 9 variables.

\section{Quantum computing for arbitrage}\label{sec:quantum_arbitrage}

As seen in the previous section, quantum computing makes use of a hybrid quantum-classical algorithm, VQE, to solve optimization problems. If we recall, three elements are needed to apply VQE: the Ising Hamiltonian, a classical optimizer and the ansatz.  

First, we need to calculate the Ising Hamiltonian, for this, the QUBO model is used.  As we already know, the Hamiltonian is expressed as the Pauli gate decomposition, (\ref{Hamiltonian}) shows a part of the Hamiltonian of the problem. As we can see, it uses I and Z gates.

\begin{equation} \label{Hamiltonian}
\begin{aligned}
    - 1.395 \cdot IIIIIIIIZ \cr
    - 1.539 \cdot IIIIIZIII \cr
    - 1.742 \cdot IIIIZIIII \cr 
    - 2.089 \cdot IIIIIIIZI \cr
    ...
\end{aligned}   
\end{equation}

One of the most used classical optimizers in quantum computing due to their good results is the Constrained Optimization by Linear Approximation, COBYLA. This optimizer is suitable when noise is not present in the cost function evaluation and it also performs only one evaluation of the objective function per optimization iteration (the number of evaluations is independent of the cardinality of the parameter set) \cite{powell_direct_1994}.

To solve the arbitrage problem using VQE in conjunction with COBYLA, the ansatz quantum circuit must be established. As we saw in the previous section, the ansatz has two hyper-parameters, entanglement and repetitions. Therefore, different combinations of these hyper-parameters have been tested in search of relevant conclusions.

As we can see in Table \ref{hyperparameters} the configuration of the ansatz entanglement that has managed to reach the optimal solution regardless of the number of repetitions has been SCA. However, using three repetitions, the optimal solution is also achieved in all the entanglements.

\begin{table}[!htb]
    \centering
    \caption{Hyper parameters that reached the optimal solution.}
    \label{hyperparameters}
    \begin{tabular}{|c|c|c|c|c|}
        \hline
        & \multicolumn{4}{c}{Entanglement} \\ 
        \hline
        Reps & Full & Linear & Circular & SCA  \\ 
        \hline
        1 & $\times$  & $\checkmark$ & $\times$  & $\checkmark$  \\
        2 & $\times$  & $\times$  & $\times$  & $\checkmark$  \\
        3 & $\checkmark$  & $\checkmark$  & $\checkmark$  & $\checkmark$  \\
        \hline
    \end{tabular}
\end{table}

\begin{figure}[h]
  \centering
  \includegraphics[width=0.8\textwidth]{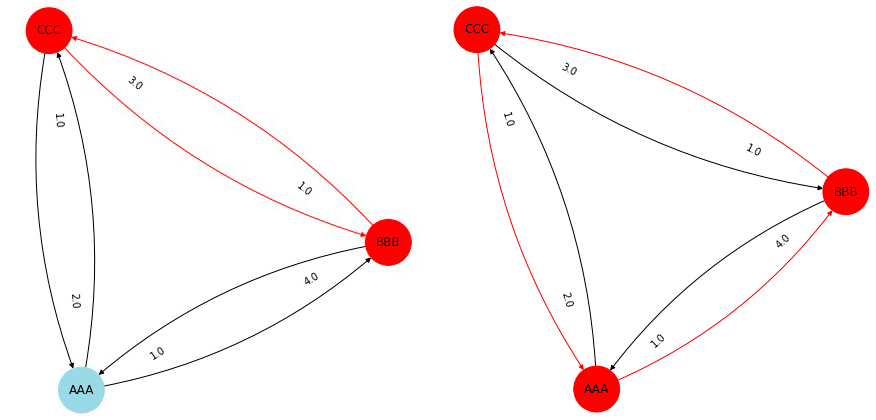}
  \caption{Comparation of a local and a global solution.}
  \label{fig:local_global}
\end{figure}

In the entanglements where the optimal solution has not been reached, the algorithm has found a solution that does not correspond to the global maximum of the problem. Fig. \ref{fig:local_global} shows a comparison of a local arbitrage solution with a 1.99 profit rate achieved by an ansatz using two repetitions and full entanglement versus the optimal solution where a gain of 23 is achieved. The main difference is that, although the local solution is feasible, it involves only two currencies. 

During the rest of this study, we will use SCA ansatz with only one repetition to maintain the depth of the circuits at the minumum for best performance on real devices. Under this circunstances, as you can see in \ref{fig:ansatz_types}, the SCA and the Circular are the same.

\section{Differential Evolution Global Optimization}\label{sec:global_de}

Although finding local solutions is feasible, when it comes to arbitrage we want to find the global solution that brings us the greatest benefit. As the number of crypto-currencies increases and, therefore, the complexity of the problem, a multitude of local solutions can arise and in the case of COBYLA we will not always find the optimal solution, but given the multitude of possibilities it can be halfway. 

\begin{figure}[h]
  \centering
  \includegraphics[width=0.6\textwidth]{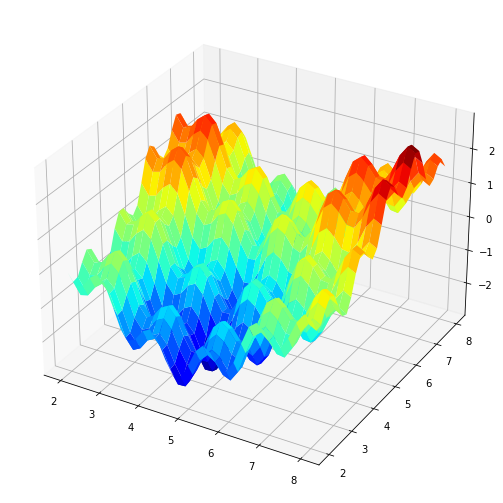}
  \caption{Global solution scenario.}
  \label{fig:global_scenario}
\end{figure}

Fig. \ref{fig:global_scenario} shows a possible inverse scenario of this type of problem. As can be seen, there are a multitude of local minimums, increasing the likelihood that the algorithm will stay with one of them. We want to achieve the global one and for this reason, a new algorithm is proposed to try to find it.

Differential evolution is an evolutionary global optimization algorithm, quite related with the genetic algorithm, that does not use gradient information and is therefore well suited to nonlinear non-differential objective functions. It is an algorithm whose operation is based on maintaining a population of candidate solutions represented as real-valued vectors. New candidate solutions are created by making modified versions of existing solutions that then replace a large part of the population each time the algorithm makes a new interaction, i.e., the base solutions are replaced by their children if the children have a better evaluation of the objective function. 

This algorithm has a number of hyper-parameters such as \emph{strategy}, which controls the type of differential evolution search that is performed. We will set a \emph{best1bin} search which involves creating new candidate solutions by selecting random solutions from the population, subtracting one from the other and adding a scaled version of the difference to the best candidate solution in the population. There are also other important parameters such as \emph{popsize}, which controls the size or number of candidate solutions that are kept in the population. The \emph{tolerance} can also be adjusted, so the lower this value, the less permissive the algorithm will be \cite{storn_differential_1997}.

\begin{algorithm}[H]
\caption{Differential Evolution Minimum Eigensolver.}\label{alg:deVQE}
    \begin{algorithmic}[1]
        \Require An operator representing the problem, an ansatz (parametrized circuit) to represent each individual, a number of individuals n, a Qiskit Runtime Session.
        \Ensure An eigenstate of the operator.
        \State Create a initial generation of n ansattzs (random parameters).
        \State Use Qiskit Runtime Estimator to calculate the expectation values of the operator for all individuals in the generation within a Session.
        \State apply scipy.optimize.differential\_evolution to calculate the next generation and iterate.
        \State Once reached convergence, use Qiskit Runtime Sampler to evaluate the ansatz with the selected parameters.
        \State Return the eigenstate.
    \end{algorithmic}
\end{algorithm}

We will use differential evolution as an optimizer to apply VQE to the arbitrage problem. However, Qiskit does not have a default function for applying differential evolution. Therefore, as described in Algorithm \ref{alg:deVQE}, a function has been created from scratch to be able to apply this method. The optimal solution with this optimizer was reached with a tolerance of 0.001, population size 15 and using 5 interactions at most. Likewise, an ansatz with 1 repetition and SCA type entanglement was used. 

\begin{figure}[h]
  \centering
  \includegraphics[width=\textwidth]{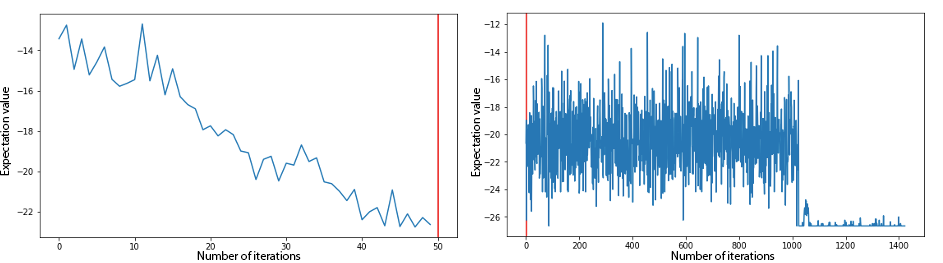}
  \caption{Expected value for individual circuit evaluations during the optimization. COBYLA (left), Differential Evolution(right).}
  \label{fig:cobyla_de}
\end{figure}

Both COBYLA and the differential evolution optimizer managed to reach the global optimum, however there is a substantial difference between them. Both seek to minimize the expected value of the Hamiltonian since both are optimizers that are used in conjunction with VQE. However, as can be seen in Fig. \ref{fig:cobyla_de}, COBYLA has a behavior similar to gradient descent, when the algorithm finds a minimum it tries to look for solutions around that minimum. On the other hand, the differential evolution tries different candidate solutions by modifying the population through the popsize parameter, that is why the expected value does not have such a marked decrease but fluctuates when calculated with new candidates.

\section{Arbitrage with three crypto-currencies}\label{sec:arbitrage_3}

As shown in \ref{sec:quantum_arbitrage}, we have used the 3 currency scenario in simulators to make the hyperparameter search for the most interesting combination to test on real hardware.

For execution on real hardware we have selected 1 repetition circular entangled ansatz due to its best fit on the topology of the IBM real quantum computers \cite{carrascal_de_las_heras_backtesting_2023}.

The main result was a 3 currency scenario that required an ansatz of 9 qubits, executed on "ibm\_geneva", a Falcon r4 QPU with 27 qubits, and quantum volume of 32:

\begin{itemize}
    \item Basis gates: CX, ID, RZ, SX, X
    \item Largest Pauli Error: 9.966e-4
    \item Largest CNOT Error: 7.725e-3
    \item Median T1: 353.86  $\mu s$
    \item Median T2: 197.83 $\mu s$
\end{itemize}

Diferential evolution algorithm executes batches of 64 circuits with 4000 shots each in an average of 104 seconds for each generation. Convergence to the global minimum was reached in 417 steps with a total time of 12 hours.

In Table \ref{estadisticos_3_currencies} there is a summary of the results obtained with 27 qubit machines on 3 asset problem.

\begin{table}[!htb]
    \centering
    \caption{\centering Three currencies on 27 qubit machine.}
    \label{estadisticos_3_currencies}
    \begin{tabular}{|c|c|c|c|c|}
        \hline
        name & QPU & circuits/batch & time(sec)/ batch \\
        \hline
        ibm\_geneva & Falcon r4 & 64 & 104 \\
        \hline
    \end{tabular}
\end{table}

\section{Arbitrage with four crypto-currencies}\label{sec:arbitrage_4}

\begin{figure}[h]
  \centering
  \includegraphics[width=0.7\textwidth]{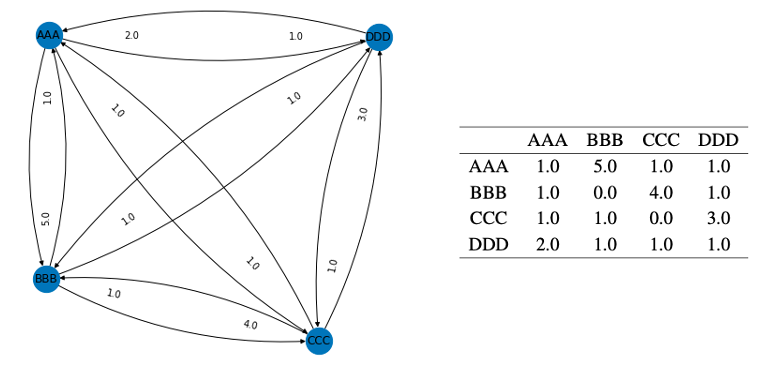}
  \caption{Exchange rates between four crypto-currencies and the transit price matrix.}
  \label{fig:scenario_4}
\end{figure}

Analogous to the arbitrage problem with three crypto-assets, a problem involving four crypto-assets is proposed to test its scalability. In this case, Fig. \ref{fig:scenario_4} shows the exchange rate graph and the price transition matrix with a clear arbitrage opportunity. The classical solution of this problem yielded a result with a profit rate of 118.99, which is quite an optimistic result which involved all four crypt-ocurrencies. 

To solve the problem in quantum form it is necessary to follow the steps detailed in the previous section, i.e., convert the problem to QUBO, create the Hamiltonian and configure the ansatz. However, in this case we will use VQE with the differential evolution optimizer due to the good results obtained in the model with three assets and to try to find the global maximum. 

\begin{figure}[h]
  \centering
  \includegraphics[width=\textwidth]{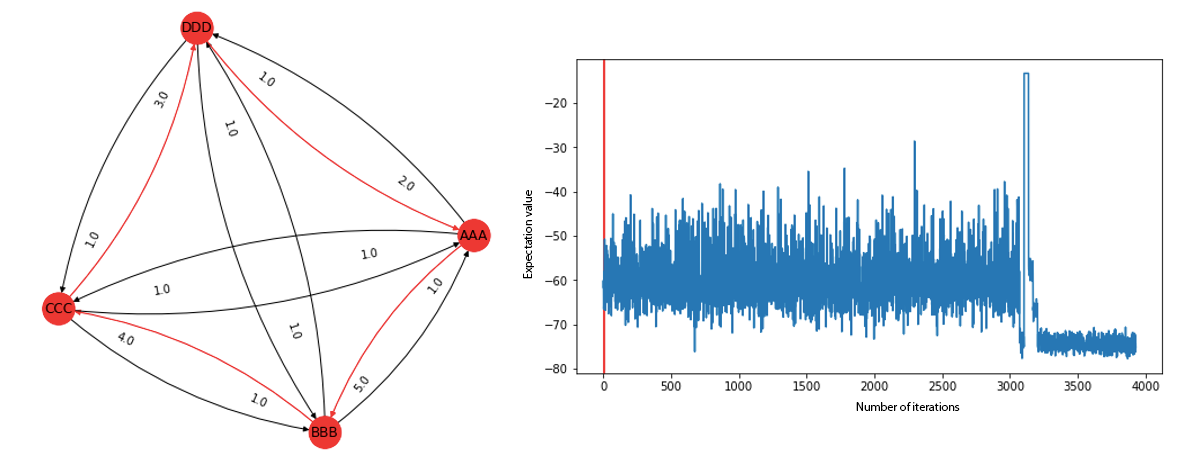}
  \caption{Differential evolution solution with four crypto-currencies.}
  \label{fig:sol_scenario_4}
\end{figure}

Fig. \ref{fig:sol_scenario_4} shows the optimal solution and the expected value plot of the differential evolution optimizer. It can be seen that the algorithm takes more iterations and thus, more time to get the minimal expected value. 

To reach this solution, the ansatz configuration was set to one iteration with SCA interleaving. This gives us the idea of the importance of choosing both the hyper-parameters of the ansatz and of the optimizer itself. To get the best suitale parameters we have performed an extensive hyperparameter search and validations.

\section{Real-data application with five crypto-currencies}\label{sec:arbitrage_5}

In this section we will present a case of real intra-exchange crypto-currency arbitrage with five assets. One of the main complications of using real data is that, given the variety of crypto-currencies and the great inequality in price and quote of them, the optimization problem presents a transition matrix with exchange rates very close to zero, which makes it difficult to find a solution. 

For this reason, in order to tackle this problem it is necessary to normalize the transition matrix. Starting from the idea that from each crypto-currency we can buy even cents, a vector with the normalization coefficients of each asset is established.

\begin{figure}[h]
  \centering
  \includegraphics[width=\textwidth]{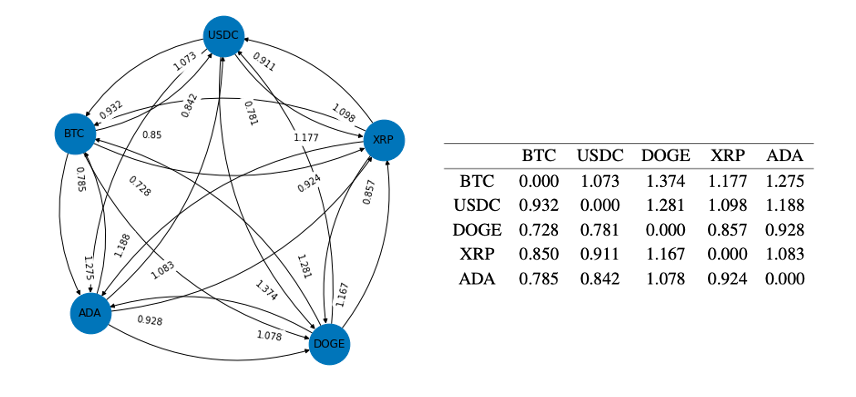}
  \caption{Transition matrix and graph of the crypto-currency of five assets.}
  \label{fig:scenario_5}
\end{figure}

In this way, we get the transition matrix of Fig. \ref{fig:scenario_5}.

The procedure to solve this problem is analogous to the previous cases, i.e., the problem needs to be transformed to QUBO to apply VQE with the differential evolution optimizer.

Due to the multitude of variables, more qubits are needed to solve the problem, specifically 25. 

\subsection{Qiskit Runtime on IBM cloud "simulator\_mps"}

\begin{figure}[h]
  \centering
  \includegraphics[width=0.8\textwidth]{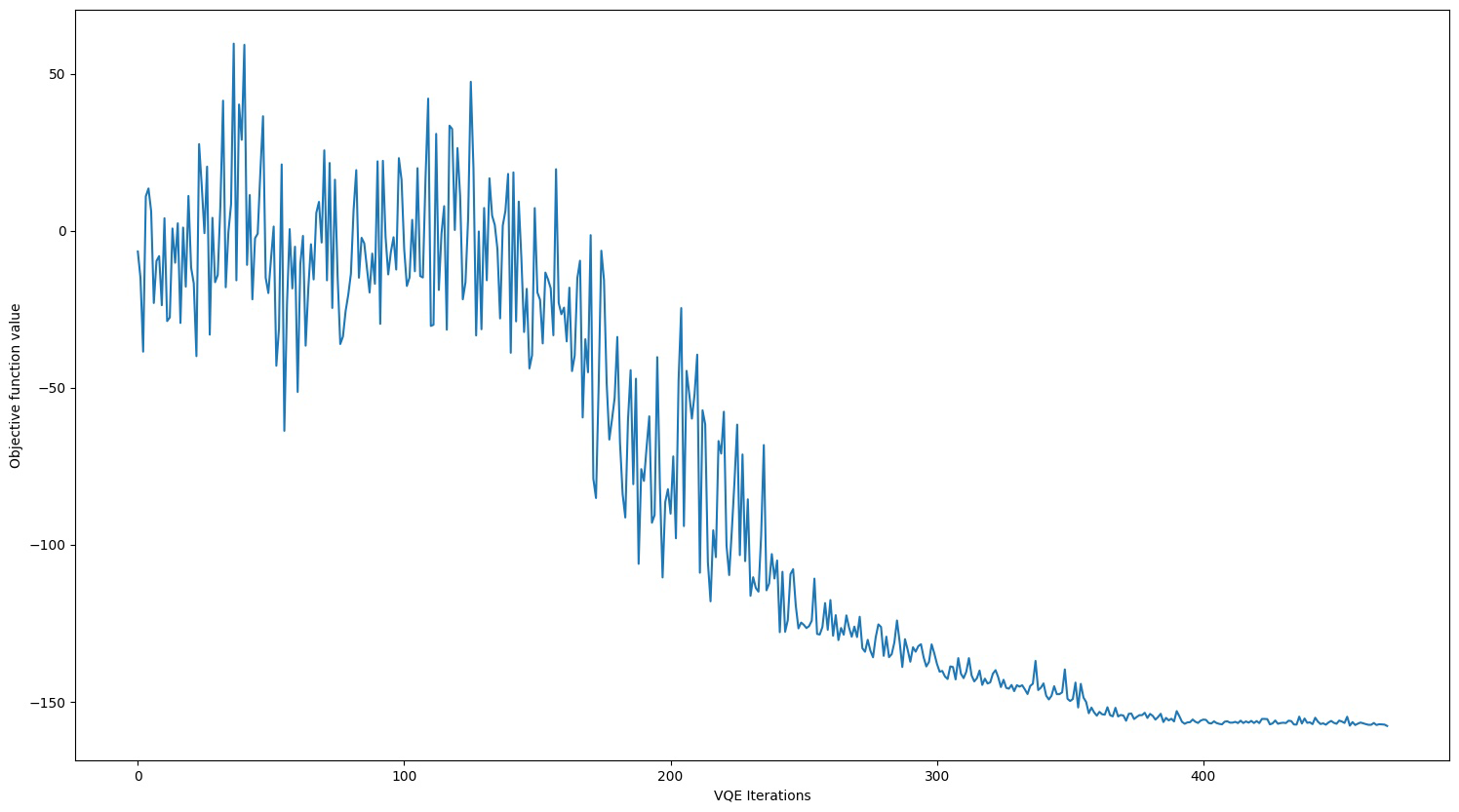}
  \caption{Simulator convergence in 5 real currencies.}
  \label{fig:convergence_5}
\end{figure}

We run a simulation using Qiskit Runtime on the IBM cloud "simulator\_mps".
Using the differential evolution optimizer, we send batches of 256 parametric circuits to the Qiskit runtime, with an average time of execution of 6 minutes. It converges after 469 steps of differential evolution (Fig. \ref{fig:convergence_5}), taking 47 hours overall. 

\begin{figure}[h]
  \centering
  \includegraphics[width=0.45\textwidth]{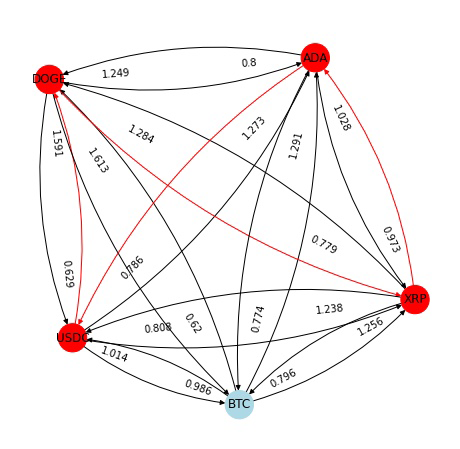}
  \caption{Transition matrix and graph of the crypto-currency of five assets.}
  \label{fig:sol_scenario_5}
\end{figure}

The solution obtained is shown in Fig. \ref{fig:sol_scenario_5}

\subsection{"ibm\_hanoi", a Falcon r5.11}

To compare the performance with the 27 qubit real machines, we have executed the 5 currency scenario on "ibm\_hanoi", a Falcon r5.11 QPU with 27 qubits, a quantum volume of 64 and 2.3K CLOPS:

\begin{itemize}
    \item Basis gates: CX, ID, RZ, SX, X
    \item Median CX Error: 9.232e-3
    \item Median SX Error: 1.977e-4 
    \item Median Readout Error: 1.250e-2
    \item Median T1: 157.54 $\mu s$
    \item Median T2: 158.15 $\mu s$
\end{itemize}

This machine executed 256 circuits with 4000 shots each in 525 seconds. A limited number of steps was made to be able to measure times. This experiment was stopped before reaching convergence in less then 24 hours.

\subsection{"ibm\_washington" Eagle r1}

To evaluate the applicability of the methodology, we have executed diferential evolution on real 127 qubit machines.
Executing in a bigger machine enables to find a best suitable path in the chip to map the circuit in an efficient way, avoiding swap gates and obtaining a shorter circuits, and thus less acumulated noise. On the other hand, bigger machines possess less CLOPS, so the execution time is longer.

\begin{figure}[h]
  \centering
  \includegraphics[width=\textwidth]{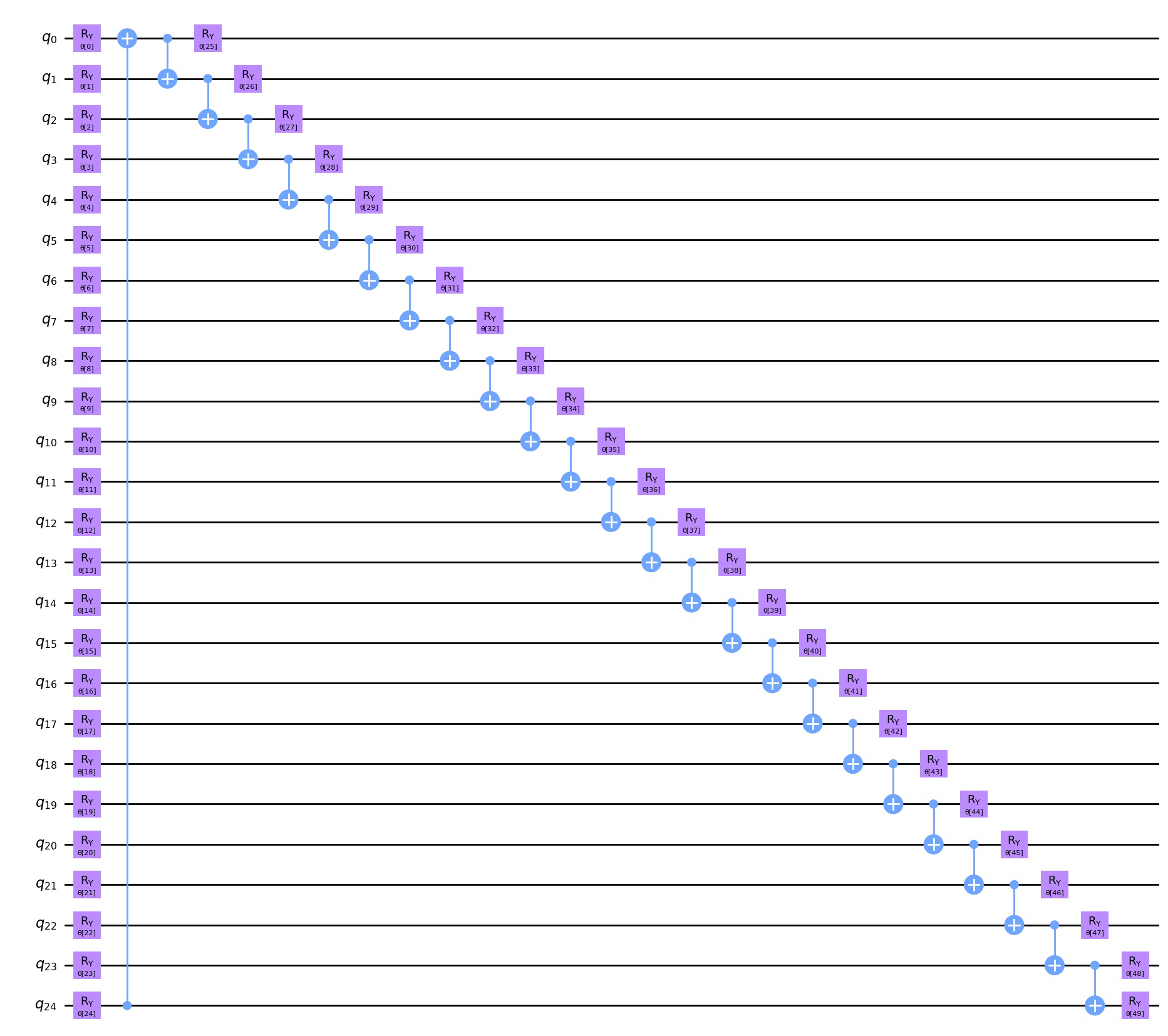}
  \caption{Ansatz for the 5 currencies scenario (25 qubits, circular entanglement).}
  \label{fig:ansatz_scenario_5}
\end{figure}

We have tested the 5 real currencies scenario in two different generations of the IBM Eagle processor. In Fig. \ref{fig:ansatz_scenario_5} we can see the 25 qubit ansatz used for this experiments.

The "ibm\_washington" ia a first generation Eagle, with Quantum Volume of 64 and CLOPS of 850:

\begin{itemize}
    \item Basis gates: CX, ID, RZ, SX, X
    \item Median CX Error: 1.258e-2
    \item Median SX Error: 2.870e-4
    \item Median Readout Error: 1.260e-2
    \item Median T1: 96.7 $\mu s$
    \item Median T2: 82.25 $\mu s$
\end{itemize}

For each step, it ran 256 circuits with 4000 shots each run in 1147 seconds. The experiment was stopped before reaching convergence in less then 24 hours.

\subsection{"ibm\_brisbane" Eagle r3}

TThird generation Eagle. The novelty in this machine is the change in the basis gates, implementing the less noisy two qubit gate ECR instead of CX.

\begin{itemize}
    \item Basis gates: ECR, ID, RZ, SX, X
    \item Median ECR Error: 7.601e-3
    \item Median SX Error: 2.184e-4
    \item Median Readout Error: 1.360e-2
    \item Median T1: 242.56 $\mu s$
    \item Median T2: 133.59 $\mu s$
\end{itemize}

It was able to ran 256 circuits with 4000 shots each run in 926 seconds. As the previous ones, this experiment was also stopped before reaching convergence in less then 24 hours.

\begin{table}[!htb]
    \centering
    \caption{\centering Executions for five currencies on real quantum machines.}
    \label{estadisticos_127_ejecuciones}
    \begin{tabular}{|c|c|c|c|c|}
        \hline
         name & QPU & circuits/batch & entanglement & time(sec)/batch \\
        \hline
        ibm\_hanoi & Falcon r5.11 & 256 & linear & 525   \\
        \hline
        ibm\_brisbane & Eagle r3 & 256 & circular & 926   \\
        \hline
        ibm\_washington & Eagle r1 & 256 & circular & 1147  \\
        
        \hline
    \end{tabular}
\end{table}

In Table \ref{estadisticos_127_ejecuciones} there is a summary of the results regarding the time needed per differential evolution generation (batch) obtained with the real quantum machines.

It is important to notice that we have achieved convergence and found the global minimun using a simulator. In real machines we have made some test to validate the viability of the methodology, but based on the number of generations in the simulator and the time taken on real computers, in these current machines the algoritmn will probably need several days of computation, which is too time consuming for this use case. These experiments show the importance of parallelization on the future of quantum computing.

The only related study we have found about applying quantum computing to arbitrage is the study from Wang et al \cite{wang_optimal_2022}. Proposes quantum algorithms for high-frequency statistical arbitrage trading, which is a quantitative trading strategy that exploits price differences between correlated assets. The quantum algorithms use variable time condition number estimation and quantum linear regression to reduce the complexity of finding cointegrated pairs of stocks. That is a complete different approach to the arbitrage problem. In our work we study the overall landscape of posible exchange paths to find arbitrage opportunities in long chains of exchanged assets instead of instant diferences between two assets.

\section{Conclusion}\label{sec:conclusion}

Quantum computing emerges as a new technology complementary to classical computing to address complex problems, with a special focus on financial problems such as arbitrage. In a world where digital assets represent the future, crypto-currency arbitrage arises as an optimization problem.

In the present work, we have tried to address crypto-currency arbitrage by proposing a quantum-classical approach by means of hybrid algorithms such as VQE. For this purpose, the scalability of the problem has been tested starting initially with three assets and finally including up to five crypto-currencies. The proposed model has been converted to QUBO format to make use of VQE and, several optimizers such as COBYLA and Differential Evolution together with a suitable configuration of the ansatz quantum circuit, have been tested in search of the global maximum of the problem.

Differential evolution has pros and cons. We have demonstrated that it has the ability to reach to the global minima, but to do so it has to pay the increase in the number of executions of quantum circuits. A interesting property of this algorithm is that all the circuits for a generation can be run in parallel. This could be a futute advantage when multiple parallel quantum systems will be available \cite{ibm_ibm_2023}.

It is precisely the scalability of the problem that means that adding more assets, increases the complexity of the problem. Despite the good results for three assets, for the rest of the cases the importance of the hyper-parameters adjustment is highlighted in order to reach the global solution. Moreover, due to the inequalities of exchange rates in the crypto-currency field, solving a problem with real data requires normalization of the exchange rate transition matrix to obtain accurate results.

As quantum computing evolves and begins to tackle practical problems, we must pay greater attention to how much work quantum computing systems can do in a given unit of time. Increased quantum processor speed is critical to support near-term algorithms based on the variational method, which requires thousands of iterations.

Nevertheless, it is a growing technology and especially in the field of crypto-currency arbitrage there is still a lot of research to be done. As future lines of research we are working in novel ways of encoding the problem to the Hamiltonian in order to be able to represent more QUBO variables using less qubits. Also a fine tuning in the multiple hyperparameters of the differential evolution has to be done to enhance the convergence and reduce the number of generations needed to achieve it.

\section*{Acknowledgment}
This work was supported by grant PID2021-123041OB-I00 funded by MCIN/AEI/ 10.13039/501100011033 and by “ERDF A way of making Europe” and by the CM under grant S2018/TCS-4423.

We acknowledge the use of IBM Quantum services for this work. The views expressed are those of the authors and do not reflect the official policy or position of IBM or the IBM Quantum team.

\bibliographystyle{plain}
\bibliography{refs}

\end{document}